\documentclass[a4paper]{revtex4}
\usepackage[english]{babel}
\usepackage{array,booktabs}
\usepackage{array} 
\usepackage{lipsum}   
\usepackage{calc}
\usepackage{pdflscape}
\usepackage{color}
\usepackage{float}
\usepackage[font=small,labelfont=bf]{caption}
\usepackage{graphicx}
\usepackage{amsmath}
\usepackage{amssymb}
\usepackage{tikz}
\usepackage{graphicx}
\usepackage{subfigure}
\usepackage{bbold}
\usepackage{textcomp}
\usepackage[nodisplayskipstretch]{setspace}

\usepackage{setspace}
\usepackage{tabularx}

\usepackage{float}
\usepackage{color}
\usepackage{amsmath}
\usepackage{float}
\usepackage{calc}
\usepackage{pdflscape}
\usepackage{color}
\usepackage{float}
\usepackage{subfigure}
\usepackage[font=small,labelfont=bf]{caption}
\usepackage{graphicx}
\begin{document}{\setlength\abovedisplayskip{4pt}}

\title{Effects Of leptonic non-unitarity on lepton flavor violation, neutrino oscillation, leptogenesis and lightest neutrino mass\footnote{
Talk presented at the XXII DAE BRNS High energy physics symposium, 12 $ - $ 16 December, 2016, Delhi University, India}
}

\author{Gayatri Ghosh$^{\star}$
}
\author{Kalpana Bora$^{\dagger}$}
\address{Department Of Physics, Gauhati University \\ Guwahati-781014, Assam, India.
\\
$^{\star}$gayatrighsh@gmail.com,$^{\dagger}$kalpana.bora@gmail.com}

\begin{abstract}
Neutrino Physics is a mature branch of science with all the three neutrino mixing angles and two mass squared differences determined with high precision. Inspite of several experimental verifications of neutrino oscillations and precise measurements of two mass
squared differences and the three mixing angles, the unitarity of the leptonic mixing matrix is not yet established, leaving room for the presence of small non-unitarity effects. Deriving the bounds on these non-unitarity parameters from existing experimental constraints, on cLFV decays such as, $ \mu\rightarrow e\gamma $,  $ \mu\rightarrow \tau\gamma $, $ \tau\rightarrow e\gamma $, we study their effects on the generation of baryon asymmetry through leptogenesis and neutrino oscillation probabilities. We consider a model where see-saw is extended by an additional singlet $ S $ which is very light, but can give rise to non-unitarity effects without affecting the form on see-saw formula. We do a parameter scan of a minimal see-saw model in a type I see-saw framework satisfying the Planck data on baryon to photon ratio of the Universe, which lies in the interval, $5.8\times 10^ {-10} < Y _{B} < 6.6 \times 10^ {-10} (BBN)$. We predict values of lightest neutrino mass, and Dirac and Majorana CP-violating phase $ \delta_{CP} $, $ \alpha $ and $ \beta $, for normal hierarchy and inverted hierarchy for one flavor leptogenesis. It is worth mentioning that all these four quantities are unknown yet, and future experiments will be measuring them.
\keywords{Neutrino mixing; flavored leptogenesis; cLFV; lepton flavor violation; Branching Ratios; baryon asymmetry}
\end{abstract}

\maketitle
\section{Introduction}	
\setlength{\baselineskip}{13pt}
Neutrinos have non-zero masses. There are 3 known flavors of neutrinos, $\nu_{e}$, $ \nu_{\mu} $, and $ \nu_{\tau} $, each of which couples only to the charged lepton
of the same flavor. $\nu_{e}$, $ \nu_{\mu} $, and $ \nu_{\tau} $  are superpositions
of three mass eigenstates, $| \nu_{\alpha} > = \sum U^{*}_{\alpha i}| \nu_{i} > $, where $ \alpha = e, \mu, \tau $ and $ \nu_{i} $ is the neutrino of definite mass $ m_{i} $. The cosmological constraints of the sum of the $ \nu $ masses bound is $ \sum_{i} m(\nu_{i}) < 0.23 \hspace{.1cm}\text{eV}$ from CMB, Planck 2015 data (CMB15+ LRG+ lensing + $H_{0}$) \cite{An}. We note that the lepton mixing matrix $ U $ has a big mixing and we know almost nothing about the phases. The discoveries of neutrino mass and leptonic mixing have come from the observation of neutrino flavor change, $ \nu_{\alpha} \rightarrow \nu_{\beta} $. CP violation interchanges every particle in a process by its antiparticle.
This CP violation can be produced by the phase $ \delta_{CP} $ in $ U $. Neutrinos can have two types of mass term in the Lagrangian $ - $ Dirac and Majorana mass terms. To determine whether Majorana masses occur in nature, so that $ \bar{\nu_{i}} = \nu_{i} $, the favorable approach to seek is Neutrinoless Double Beta Decay ($0\nu\beta\beta$). 
\par
In the conventional type I see-saw framework, there are Dirac and Majorana mass matrices $m_{D}$ and $M_{R}$ in the Lagrangian,
\begin{equation}
L = \frac{1}{2}\bar{N_{R}}M_{R}N_{R}^{c} + N_{R}m_{D}\nu_{L} +h.c.
\end{equation}
The low energy mass matrix is given by 
\begin{equation}
m_{\nu} = -m_{D}^{T}M_{R}^{-1}m_{D}.
\end{equation}
In the usual unitarity scenario, the three active neutrinos, the flavor eigen states $ \nu_{e} $, $ \nu_{\mu} $, $ \nu_{\tau} $ are connected to the mass eigen states $ \nu_{1} $, $ \nu_{2} $, $ \nu_{3} $ via $ \nu_{\alpha}  = N_{\alpha i} \nu _{i} $, where $ N^{\dagger} N= 1 $. Here $ N $ is the generalized neutrino mixing matrix which could be either unitary or non-unitary. In the diagonal charged lepton basis, $ m_{\nu} $ is diagonalised by a unitary matrix as 
\begin{equation}
UPm_{\nu}P^{\dagger}U^{\dagger} = m_{\nu}^{D}.
\end{equation}
The Pontecorvo-Maki-Nakagawa-Sakata (PMNS) matrix is UP, where U is 
\begin{equation}
U = \begin{pmatrix}
c_{12}c_{13} & s_{12}c_{13} & s_{13}e^{-i\delta}\\
-s_{12}c_{23}-c_{12}s_{23}s_{13}e^{i\delta} & c_{12}c_{23}-s_{12}s_{23}s_{13}e^{i\delta}& s_{23}c_{13}\\
s_{12}s_{23}-c_{12}c_{23}s_{13}e^{i\delta} & -c_{12}s_{23}-s_{12}c_{23}s_{13}e^{i\delta} & c_{23}c_{13}\\
\end{pmatrix}.
\end{equation}
Here $\theta_{12} = 33.56^{\circ}, \theta_{23} = 41.6^{\circ} (50^{\circ}), \theta_{13} = 8.46^{\circ} (8.49^{\circ})$ \cite{Fg} (see ref. \cite{Fg} for recent global fit values) are the solar, atmospheric and reactor angles for Normal Ordering (Inverted Ordering) respectively. 
The Majorana phases reside in P, where 
\begin{equation}
P = \text{diag} \begin{pmatrix} 1 & e^{i\alpha} & e^{i(\alpha + \beta)}\end{pmatrix}.
\end{equation}
Cosmologists suggest that just after the Big Bang, the universe contained equal amounts of matter and antimatter. Today the universe contains matter but almost no antimatter. This change needs that matter and antimatter act differently (CP violation). The CP-violating scenario to explain this change is leptogenesis. Leptogenesis is a natural outcome of the see-saw Mechanism. In the see-saw picture, we assume that,
just as there are 3 light neutrinos $ \nu_{1}, \nu_{2}, \nu_{3} $,
there are 3 heavy right-handed neutrinos $ M_{1}, M_{2}, M_{3} $, where $ M_{R} \sim 10^{9-14} $ GeV, $ M_{R}\sim M_{1}, M_{2}, M_{3} $ which were there in the Hot Big Bang. The $ M_{R} $ decays modes are:
\begin{equation}
M \rightarrow l^{-} + H^{+} ,
M \rightarrow  l^{+} + H ^{-}, M \rightarrow \nu + H^{0}, M \rightarrow \bar{\nu} + \bar{H^{0}},
\end{equation} 
where, $ l^{-} $ are $ e^{-}, \mu^{-}, \tau^{-} $ and $ H^{+}, H ^{-}, H ^{0}$ are SM Higgs. CP violation effects in the $ M_{R} $ decays, may result from phases in the decay coupling constants. This leads to unequal numbers of leptons ($l^{-} \text{and} \hspace{.1cm} \nu$) and antileptons ($l^{+} \text{and} \hspace*{.1cm} \bar{\nu}$) in the Universe,
\begin{equation}
\Gamma(M \rightarrow  l^{+} + H ^{-} ) \neq  \Gamma(M \rightarrow l^{-} + H^{+}).
\end{equation}
In leptogenesis, CP-violating decays of heavy Majorana neutrinos create a lepton $-$ antilepton asymmetry \cite{ma} and then B+L violating sphaleron processes \cite{BL} at and above the electroweak symmetry breaking scale converts part of this asymmetry into the observed baryon-antibaryon asymmetry. The heavy neutrinos are see-saw partners of the observed light ones. 
\par 
Depending on mass of the lightest heavy RH Majorana neutrinos (whose decay causes leptogenesis) the leptogenesis can be of three types $-$ unflavored (or one flavor), two flavored and three flavored leptogenesis. For unflavored leptogenesis, valid for $M_{1} \geq 10^{11}$ GeV, we have taken here $M_{1} \sim 10^{12}$ GeV where the flavor of the final state leptons play no role. It can be shown that for lower values of $M_{1}$ it depends on the flavor of the final state leptons, and hence is called flavored leptogenesis \cite{fl}. Here we consider unflavored leptogenesis. For unflavored leptogenesis, the decay asymmetry in the case of hierarchical heavy neutrinos is given by \cite{Wern}
\begin{equation}
\varepsilon_{1} =\frac{1}{8\pi\upsilon^{2}}\frac{1}{(m_{D}m_{D}^{\dagger})_{11}}\sum_{2,3}Im{(m_{D}m_{D}^{\dagger})^{2}_{1j}}f(M_{j}^{2}/M_{1}^{2})
,\end{equation}
where, $f(x) = \frac{-3}{2 \sqrt{x}}$ for $ x > $ 1, i.e., for hierarchical heavy neutrinos. The baryon asymmetry of the Universe is proportional to the decay asymmetry $ \varepsilon_{1} $. The Dirac mass matrix $ m_{D} $ in terms of a complex and orthogonal matrix $R$ is \cite{Dirac}

\begin{equation}
m_{D} = i \sqrt{M_{R}}R\sqrt{m_{\nu}^{diag}}U^{\dagger}.
\end{equation}
The compatible quantity for leptogenesis is then,
\begin{equation}
m_{D}m_{D}^{\dagger} = \sqrt{M_{R}}R\sqrt{m_{\nu}^{diag}}U^{\dagger}U\sqrt{m_{\nu}^{diag}}R^{\dagger}\sqrt{M_{R}}
=\sqrt{M_{R}}Rm_{\nu}^{diag}R^{\dagger}\sqrt{M_{R}}
\end{equation}
If $R$ is real then there is no leptogenesis at all. Here, we have taken $ R = U $ (see Eq.(4)). $ R $ consists of the low energy mixing elements and the CP phases. Since unitarity of neutrino mixing matrix has not been proved yet, if it is non-unitary, then for the neutrino mixing matrix N to be non-unitary, we have
\begin{equation}
\nu_{\alpha} = N_{\alpha i}\nu_{i},
\end{equation}
connecting the flavor and mass states. The non-unitary matrix N can be written as
\begin{equation}
N = (1 + \eta)U_{0},
\end{equation}
where $ U_{0} = U*P $. If $m_{\nu}$, which is diagonalized by a non-unitary mixing matrix, originates from the see-saw mechanism, we have
\begin{equation}
m_{D} = i \sqrt{M_{R}}R\sqrt{m_{\nu}^{diag}}N^{\dagger}.
\end{equation}
And thereupon, we have
\begin{equation}
m_{D}m_{D}^{\dagger} = \sqrt{M_{R}}R\sqrt{m_{\nu}^{diag}}N^{\dagger}N\sqrt{m_{\nu}^{diag}}R^{\dagger}\sqrt{M_{R}}
.\end{equation}
For non-unitary $ U_{PMNS} $ matrix, $ N^{\dagger}N = 1 + 2 U_{0}^{\dagger}\eta U_{0} \neq 1$. Leptogenesis is no longer independent of the low-energy phases. It depends on the phases in $ U_{0} $ as well as to the phases in $ \eta $. Leptogenesis \cite{le} is one of the exceedingly well-inspired framework which produces baryon asymmetry of the Universe through B + L violating electroweak sphaleron process \cite{BL,sa}. 
\par 
In the conventional type I see-saw mechanism, due to the mixing of the left handed and right-handed neutrinos the PMNS matrix is non-unitary. Nevertheless this non-unitarity is too small to have any observable effects in Lepton Flavor Violation (LFV) or neutrino oscillation, (see Ref. \cite{Wern, pil,frt}). If we want to connect the lepton mixing matrix to leptogenesis via the Casas-Ibarra parametrization, then there should not be any significant sizeable contribution to $ m_{\nu} $ and leptogenesis other than
the usual see-saw terms. Hence we must decouple the origin of unitary violation from these terms. Mixing of the light neutrinos with new physics, creates non-unitarity in the low-energy mixing matrix. We consider here a model as used in Ref. \cite{Wern} where the see-saw mechanism is enlarged by an additional singlet sector, which leads to a  9 $\times$ 9 mass matrix. Here
\begin{equation}
L = \frac{1}{2}\begin{pmatrix}
\bar{\nu_{L}^{c}} & \bar{N_{R}}& \bar{X}\\
\end{pmatrix}\begin{pmatrix}
0 & m_{D}^{T} & m^{T}\\
m_{D} & M_{R} & 0\\
m & 0 & M_{s}\\
\end{pmatrix} \begin{pmatrix}
\nu_{L}\\
N_{R}^{c}\\
X^{c}\\
\end{pmatrix},
\end{equation}
where the upper left block is the one corresponding to usual type I see-saw mechanism. It can be diagonalised with a unitary matrix $ F$, such that, 
\begin{equation}
F^{T}\begin{pmatrix}
0 & m_{D}^{T} & m^{T}\\
m_{D} & M_{R} & 0\\
m & 0 & M_{s}\\
\end{pmatrix} F = \begin{pmatrix}
m_{\nu}^{diag} & 0  & 0\\
0 & M_{R} & 0\\
0 & 0 & M_{s}^{diag}\\
\end{pmatrix}.
\end{equation}
The form of $F$ is as given in Ref. \cite{Wern}.
As discussed in Ref. \cite{Wern} with proper choice of various elements in $ F $ and of $ M_{s}$, it can be shown that
\begin{equation}
m_{\nu}^{diag} = -\tilde{N^{T}}m_{\nu}^{D}M_{R}^{-1}m_{\nu}^{D}\tilde{N},
\end{equation}
which shows that $N = (\tilde{N^{\dagger}})^{-1}$ is the lepton mixing matrix, as there is no other significant contribution to the mass term of the light neutrinos. Thus in this model, the usual see-saw mechanism remains unaltered with unmodified leptogenesis. Here $ M_{R} $ does not couple to the new singlets but a sizeable non-unitary lepton mixing matrix $ N $ can be induced, thus providing us with a framework where we can apply Eq.(14). 
\par
We consider here, that lepton asymmetry is generated by out of equilibrium decay of heavy right-handed Majorana neutrinos into Higgs and lepton within the framework of type I see-saw mechanism. In a hierarchical case of three right-handed heavy Majorana neutrinos $M_{2,3} > M_{1}$,  the lepton asymmetry created by the decay of $ M_{1} $, the
lightest of three heavy right-handed neutrinos is \cite{fla} 
\begin{eqnarray}
\varepsilon_{1}^{\alpha} & = \frac{1}{8\pi\upsilon^{2}}\frac{1}{(m_{D}^{\dagger}m_{D})_{11}}\biggl[\sum_{2,3}Im[{(m_{D}^{*})_{\alpha 1}(m_{D}^{\dagger}m_{D})_{1j}(m_{D}^{*})_{\alpha j}}]g(x_{j}) \nonumber  \\ & + \sum_{2,3}Im[{(m_{D}^{*})_{\alpha 1}(m_{D}^{\dagger}m_{D})_{j1}(m_{D}^{*})_{\alpha j}}]\frac{1}{1-x_{j}}\biggr].
\end{eqnarray}

Here $ \upsilon $ = 174 GeV, is the vacuum expectation value of the SM Higgs and,
\begin{equation}
g(x) = \sqrt{x}(1 + \frac{1}{1-x} - (1+x)ln\frac{1+x}{x}), x_{j} = \frac{M_{j}^{2}}{M_{1}^{2}}.
\end{equation}
At temperatures, $T \geq 10^{12}$ GeV all charged lepton flavors come out of equilibrium and thus all of them behave in the same way which results in the one flavor regime. At moderate temperatures  $T < 10^{12}$ GeV  ($T < 10^{9}$ GeV), tau (muon) Yukawa coupling interactions come into equilibrium and hence flavor effects play an important role in the calculation of lepton asymmetry \cite{a,b,c,d,e,gg}. The region of temperatures belonging to $10^{9}< T/GeV < 10^{12}$ and $T/GeV < 10^{9}$ are respectively denoted as two and three flavor regimes of leptogenesis \cite{f}.

\par 
The building blocks of matter are the quarks, the charged leptons, and the neutrinos.
The discovery and study of the Higgs boson at the Large Hadron Collider (LHC) has provided strong evidence that the quarks and charged leptons derive their masses
from a coupling to the Higgs field. Most theorists strongly believe that the origin of
the neutrino masses is different from the origin of the quark and charged lepton masses.
Neutrino oscillation has proved that neutrinos have non-zero masses. We, and all matter may have descended from heavy neutrinos. We list the values of $m_{lightest}$ for  one flavor for different hierarchies and unitarity, non-unitarity of $U_{PMNS}$  in Table II, and check whether our values of $ m_{lightest} $ are consistent with the constraints on the absolute scale of $ \nu $ masses. The new results presented in this work are as follows:
\newline
$\bullet$ We have calculated new values of non$-$unitarity parameters of $U_{PMNS}$ matrix from the bounds on rare cLFV decays.
\newline
$ \bullet$ Hence we predicted the absolute value of lightest $ \nu $ mass in this regard. The values of lightest $\nu$ mass lies in the range of 0.0018 eV to 0.0023 eV, 0.048 eV to 0.056 eV, 0.05 eV to 0.054 eV, 0.053 eV to 0.062 eV in one flavor leptogenesis regime. 
\newline
$\bullet$ All these values satisfy the constraint, $ \sum_{i} m(\nu_{i}) < 0.23 \hspace{.1cm}\text{eV}$.
\newline 
$\bullet$ The predicted values of CP-violating phases, $ \delta_{CP} $ and Majorana phases $\alpha $, $\beta $ are $36^{\circ}$, $72^{\circ}$, $108^{\circ}$, $144^{\circ}$, $180^{\circ}$, $216^{\circ}$, $252^{\circ}$, $288^{\circ}$, $324^{\circ}$, $360^{\circ}$.
\par 
The paper is organized as follows. In Section II, we show the effect of low energy phenomenology of non-unitarity on charged lepton flavor violating decays in type I see-saw theories and present the values of various parameters used in our analysis for the generation of baryon asymmetry of the Universe through the mechanism of leptogenesis. Section III contains our calculations and results. Section IV contains analysis and discussions. Section V summarizes the work.

\section{Low Energy Phenomenology of Non-Unitarity and Leptogenesis}
One interesting feature of non-unitarity of the PMNS matrix can be studied in rare charged lepton flavor violation (LFV) decay processes. In the light of unitarity violation in decays such as
$ \alpha \rightarrow \beta \gamma, (\alpha, \beta ) = (\tau, \mu), (\tau, e) \hspace{.1cm} \text{or} (\mu, e)$, the branching ratio is \cite{Wern}
\begin{equation}
\frac{BR( \alpha  \rightarrow  \beta +  \gamma )}{BR( \alpha  \rightarrow  \beta +  \nu \bar{\nu} )} = \frac{100 \alpha}{96 \pi}|(NN^{\dagger})_{\alpha \beta}|^{2}.
\end{equation}
Also 
\begin{equation}
\frac{BR( \tau  \rightarrow  \mu +  \gamma )}{BR( \tau  \rightarrow  \mu +  \nu \bar{\nu} )} \simeq \frac{25 \alpha}{6 \pi}|\eta_{\mu\tau}|^{2}; \hspace{.1cm} \frac{BR( \tau  \rightarrow  \mu +  \gamma )}{4.2\times 10^{-10}} \simeq \frac{|\eta_{\mu\tau}|^{2}}{25\times 10^{-8}}.
\end{equation}
Using the latest updated constraint on $BR( \tau  \rightarrow  \mu +  \gamma ) = 4.4 \times 10^{-8}$ \cite{baldini}, one can derive bounds on $|\eta_{\mu\tau}|$ from Eq.(22). It can be shown that
\begin{equation}
\frac{BR( \tau  \rightarrow  \mu +  \gamma )}{BR( \mu  \rightarrow  e +  \gamma )} =  \frac{BR( \tau  \rightarrow  \mu +  \nu_{\tau} \bar{\nu_{\mu}} )}{BR( \mu  \rightarrow  e +  \nu_{\mu} \bar{\nu_{e}} )} \times \frac{|\eta_{\mu\tau}|^{2}}{|\eta_{\mu e}|^{2}}.
\end{equation}
Now, we calculate the ratio, $$ \frac{BR( \tau  \rightarrow  \mu +  \nu_{\tau} \bar{\nu_{\mu}} )}{BR( \mu  \rightarrow  e +  \nu_{\mu} \bar{\nu_{e}} )} = 0.176745. $$ Thus we find constraints on $ |\eta_{\mu e}|$ from Eq.(23), using the latest constraint on $ BR( \mu  \rightarrow  e +  \gamma ) $, where $BR( \mu  \rightarrow  e +  \gamma ) = 4.2 \times 10^{-13}$ \cite{baldini}. 
Again we have
\begin{equation}
\frac{BR( \tau  \rightarrow  \mu +  \gamma )}{BR( \tau  \rightarrow  e +  \gamma )} =  \frac{BR( \tau  \rightarrow  \mu +  \nu_{\tau} \bar{\nu_{\mu}} )}{BR( \tau  \rightarrow  e +  \nu_{\mu} \bar{\nu_{e}} )} \times \frac{|\eta_{\mu\tau}|^{2}}{|\eta_{\tau e}|^{2}}.
\end{equation}
From our calculation, the ratio $\frac{BR( \tau  \rightarrow  \mu +  \nu_{\tau} \bar{\nu_{\mu}} )}{BR( \tau  \rightarrow  e +  \nu_{\mu} \bar{\nu_{e}} )}$ is, 
 \begin{equation}
 \frac{BR( \tau  \rightarrow  \mu +  \nu_{\tau} \bar{\nu_{\mu}} )}{BR( \tau  \rightarrow  e +  \nu_{\mu} \bar{\nu_{e}} )} = 2.509.
 \end{equation}
And then we calculate the latest updated bounds on $|\eta_{\tau e}|$. The calculations are summarized in Table I. We note that interesting results on cLFV $ \mu\rightarrow e \gamma $ in NUSM, NUHM, NUGM, mSUGRA models are presented in Ref. \cite{ggg} in which we have predicted some values of new SUSY particles that may be detected at next run of LHC.

\begin{table}[phtb]
\begin{center}
\begin{tabular}{|c|c|c|} \toprule
\hline
Serial No.& Latest updated Branching Ratios & Calculated bounds  \\
&  on cLFV Decays & on $ |\eta|_{\alpha\beta} $\\ \colrule
1 & $BR( \mu  \rightarrow  e +  \gamma ) = 4.2 \times 10^{-13}$  
&  $ |\eta_{\mu e}| =6.64733013 \times 10^{-6}$  \\
2 & $BR( \tau  \rightarrow  \mu +  \gamma ) = 4.4 \times 10^{-8}$  
& $ |\eta_{\tau \mu}| = 5.11766 \times 10^{-3}$  \\
3 & $ BR( \tau  \rightarrow  e +  \gamma ) = 3.3 \times 10^{-8}$
&  $ |\eta_{\tau e}| = 7.021 \times 10^{-3}$ \\
 \hline
\end{tabular} \label{ta1}
\end{center}
\caption{Our calculated constraints on non-unitarity parameter $ \eta_{\tau e}, \eta_{\tau \mu}, \eta_{\mu e} $ using branching ratios of latest cLFV decays taken from \cite{baldini}}
\end{table}

Returning to leptogenesis, the baryon asymmetry should lie in the interval,  $5.8\times 10^{-10} < Y_{B} < 6.6 \times 10^{-10} $ \cite{pdg}. In general, we have taken complex and orthogonal matrix $R = U_{PMNS}$, but R can be taken as $R = V_{CKM} \times U_{PMNS} $ in non SUSY SO(10) models \cite{gg} (see ref. \cite{gg} for detailed discussion on it) as studied in the context of breaking entanglement of octant of $ \theta_{23} $ and $ \delta_{CP} $ in the light of baryon asymmetry of the universe through the mechanism of leptogenesis.
\par 
For the Normally ordered light neutrino masses, we have 
\begin{equation}
M_{R}^{diag} = \text{diag} (M_{1} , M_{2}, M_{3} )
= M_{1} \text{diag}(1, \frac{M_{2} }{M_{1}} , \frac{M_{3}}{M_{1}} )
= M_{1}\text{diag}(1, \frac{m_{1}}{m_{2}} ,\frac{m_{1}}{m_{3}}),
\end{equation}
with $m_{1} \in [10^{-6} eV , 10^{-1} eV]$,
and, $m_{2}^{2} - m_{1}^{2} = 7.60 \times 10^{-5} eV^{2}, m_{3}^{2} - m_{1}^{2} = 2.48 \times 10^{-3} eV^{2} $ as is evident from the neutrino oscillation data \cite{Fg}, $m_{1}$ being the lightest of three neutrino masses. For the inverted ordered light neutrino masses, we have 
\begin{equation}
M_{R}^{diag} = \text{diag} (M_{1} , M_{2}, M_{3} )
= M_{1} \text{diag}(1, \frac{M_{2} }{M_{1}} , \frac{M_{3}}{M_{1}} )
= M_{1}\text{diag}(1, \frac{m_{1}*m_{3}}{m_{2}^{2}} ,\frac{m_{1}}{m_{2}}),
\end{equation}
with $ m_{3} $ being the lightest of the three neutrino masses. Next, we do the parameter scan for one flavored leptogenesis of a minimal see-saw model satisfying the Planck
data on baryon to photon ratio of the universe for four cases :
\newline
(i) Normal hierarchical structure neutrino masses, non-unitarity of PMNS matrix.\\
(ii) Normal hierarchical structure neutrino masses, unitarity of PMNS matrix.\\
(iii) Inverted hierarchical structure of neutrino masses, non-unitarity of PMNS matrix.\\
(iv) Inverted hierarchical structure neutrino masses, unitarity of PMNS matrix.
\par 
We perform random scan of the parameter space for NH, IH in the light of recent ratio of
the baryon to photon density bounds $5.8\times 10^{-10} < \eta_{B} < 6.6 \times 10^{-10} $ in the following ranges:

\begin{eqnarray}	
\begin{array}{rcl}
m_{1}(m_{3} ) & \in & [10^{-6}\hspace{.1cm} eV, 0.1 \hspace{.1cm} eV]\hspace{.1cm} ([10^{-6} \hspace{.1cm} eV, 0.1 \hspace{.1cm} eV] ),\\[8pt]
\delta_{CP} & \in & [ 0, 2\pi],\\
\alpha & \in & [ 0, 2\pi], \\
\beta & \in & [ 0, 2\pi].
\end{array}
\end{eqnarray}

While doing parameter scan, we find values of lightest neutrino mass, Majorana phases $ \alpha $, $ \beta $ and Dirac CPV phase $ \delta_{CP} $, for which baryon to photon ratio $ Y_{B} $ lies in the given range, for above four cases. This is done for one flavor/unflavored leptogenesis regime.

\section{Calculations and Results}
Results of our analysis have been presented in Fig. 1 $ - $ Fig. 9. It can be seen from Fig. 1 that in the one-flavor regime, NH structure of neutrino masses, non-unitarity textures of PMNS matrix can give rise to correct baryon asymmetry of the Universe, $5.8\times 10^{-10} < Y_{B} < 6.6 \times 10^{-10}$, if the lightest 
$\nu$ mass lies around 0.0018 eV to 0.0023 eV.  A close numerical inspection of the situation reveals that for a lightest neutrino mass of 0.0024 eV-0.005 eV, one can exceed the upper bound on $ Y_{B} $.
\begin{figure}[htb]
\centerline{\includegraphics[width= 7.8 cm]{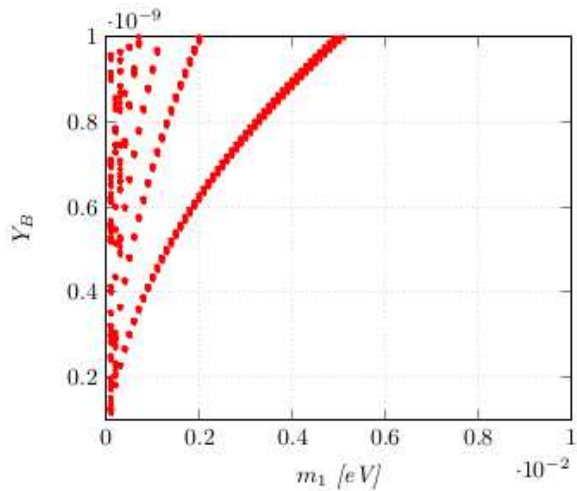}}
\caption{Scatter plot of the lightest neutrino mass $m_{1}$ against the baryon asymmetry of the Universe with normal hierarchy, non-unitarity case in one flavor leptogenesis regime. The values of $ m_{1} $ [eV] along the x-axis is multiplied by $ 10^{-2}$.}
\end{figure}
The dependence of lightest neutrino mass $ m_{1} $ on Majorana phases $\alpha$, $\beta$ is shown in the left, right panel of Fig. 2 respectively, $ Y_{B} $ being constrained in the order $10^{-10 }$. We find that $M_{1}= 10^{12}$ GeV is favoured in the light of baryon asymmetry of the Universe for one flavor regime.
\begin{figure}[htb]
\centerline{\begin{subfigure}[]{\includegraphics[width= 6.5 cm]{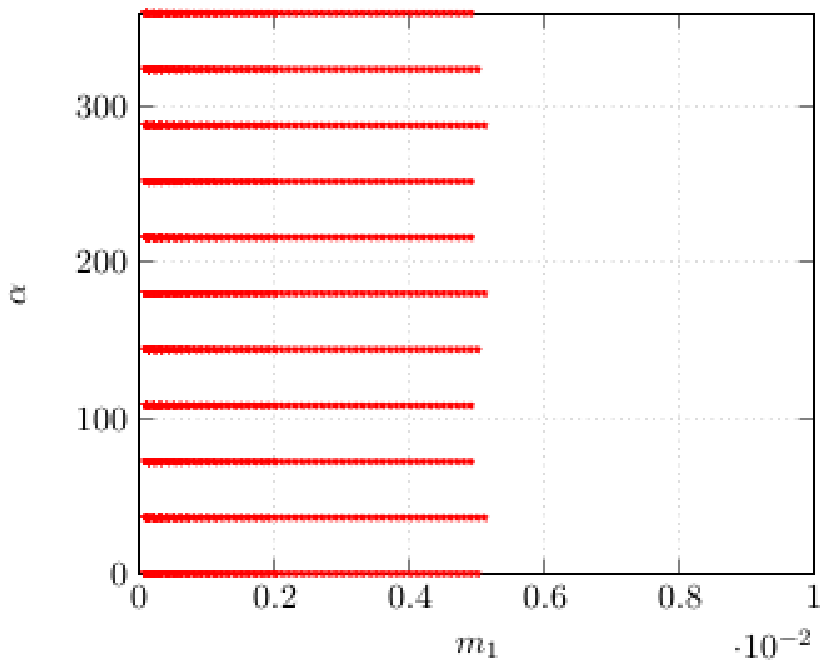}}\end{subfigure}
\begin{subfigure}[]{\includegraphics[width= 6.5 cm]{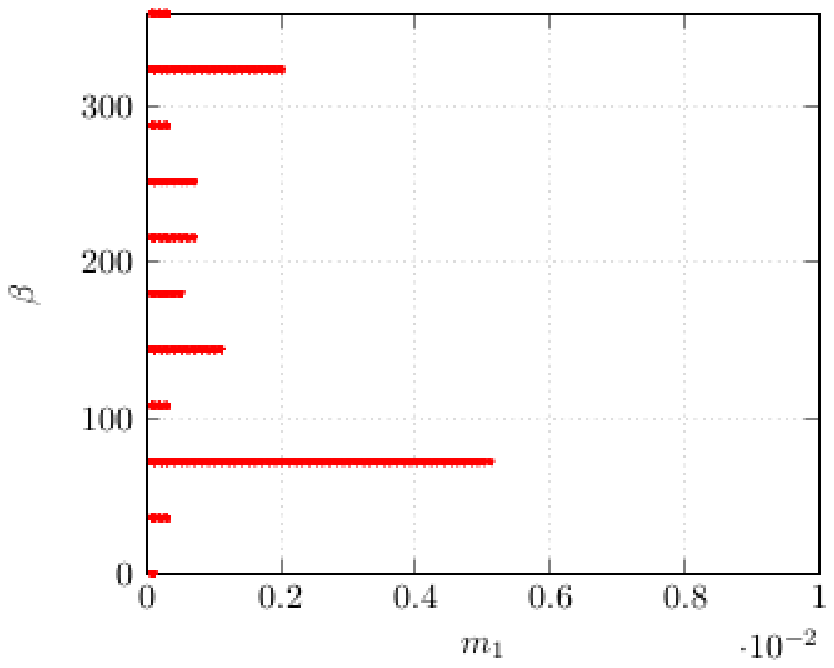}}\end{subfigure}}
\caption{Variation of lightest neutrino mass $ m_{1} $ with Majorana phases $ \alpha $ and $ \beta $ in case of NH, non-unitarity case in one flavor leptogenesis regime. $m_{1}$ is in eV. The values of $ m_{1} $ [eV] in Fig. a, b along the x-axis is multiplied by $ 10^{-2}$}
\end{figure}

Figure 3 shows the scatter plot of the lightest neutrino mass $ m_{1} $ against the baryon asymmetry of the Universe with Normal hierarchy, and unitary $U_{PMNS}$ in one flavor regime. For $Y_{B}$ to be in the range, $5.8\times 10^{-10} < \eta_{B} < 6.6 \times 10^{-10} $, $m_{1}$ lies between 0.048 eV to 0.056 eV. For $ Y_{B} $ in the order $10^{-10}$, the lightest neutrino mass $m_{1}$ is mostly concentrated in the region 0.043 eV to 0.006 eV.

\begin{figure}[tbh]
\centerline{\includegraphics[width= 7.8 cm]{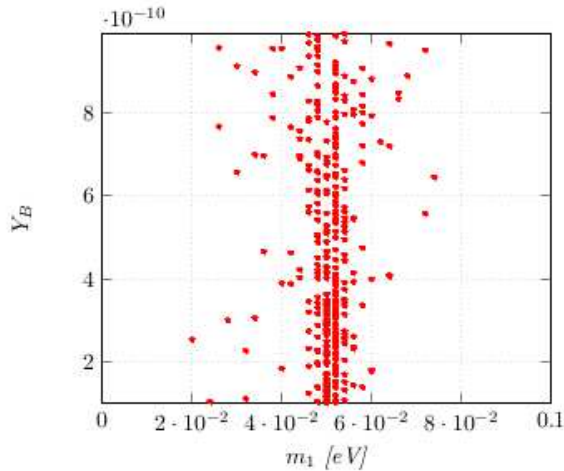}}
\caption{Scatter plot of the lightest neutrino mass $ m_{1} $
against the the baryon asymmetry of the
Universe with normal hierarchy, unitarity case
in one flavor leptogenesis regime.}
\end{figure}

In Fig. 4 we have shown the variation of lightest neutrino mass $ m_{1} $ with Dirac CP phase $\delta_{CP}$ and Majorana phase $ \alpha $ for NH (normal hierarchy), unitarity texture. For $ Y_{B} $ to be in the consistent BAU range $5.8\times 10^{-10} < Y_{B} < 6.6 \times 10^{-10} $, one of the values of $ \delta_{CP} $ predicted by us, i.e. $\delta_{CP} = 252.9^{\circ}$ is also favoured in the recent global fit values, $\delta_{CP} = 253.8^{\circ}$  for normal hierarchy.
\begin{figure}[tbh]
\centerline{\begin{subfigure}[]{\includegraphics[width= 6.5 cm]{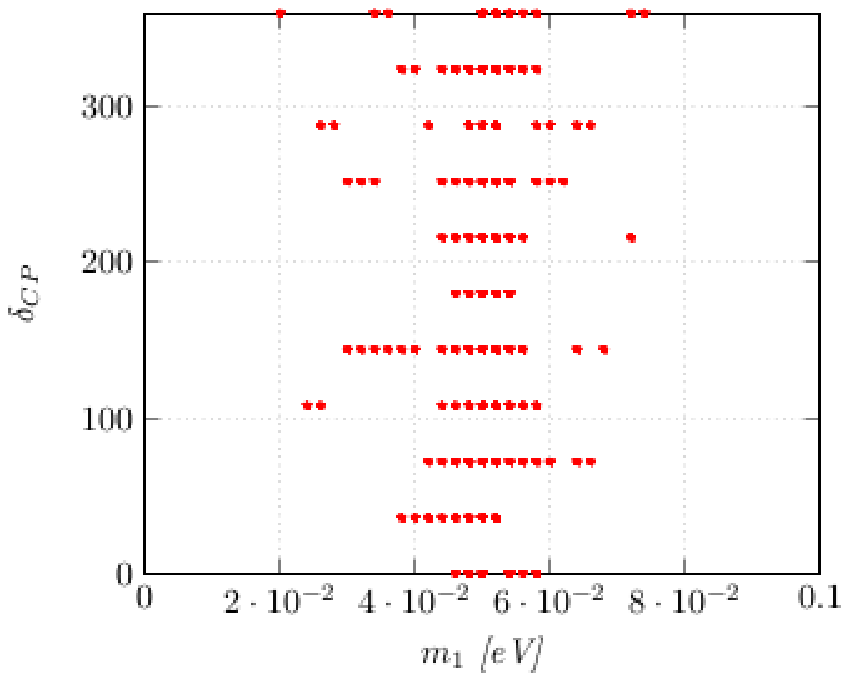}}\end{subfigure}
\begin{subfigure}[]{\includegraphics[width= 6.5 cm]{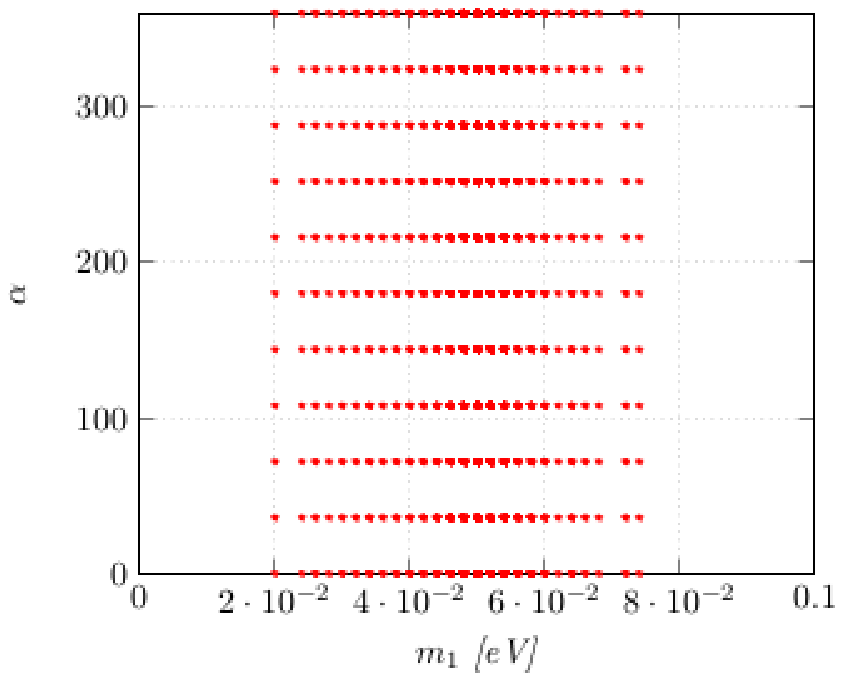}}\end{subfigure}}
\caption{Variation of lightest neutrino mass $ m_{1} $ on Dirac CP phase $ \delta_{CP} $ and Majorana phases $ \alpha $ in case of NH unitarity case in one flavor leptogenesis regime.}
\end{figure}
It can be seen from Fig. 5 that in the one-flavor regime, IH structure of neutrino masses, non-unitarity textures of PMNS matrix can give rise to baryon asymmetry of the Universe, of the order of $10^{-10}$, if the lightest neutrino mass lies around 0.05 eV to 0.054 eV. Few Points lie in the region, $5.8\times 10^{-10} < Y_{B} < 6.6 \times 10^{-10} $. 
\begin{figure}[tbh]
\centerline{\includegraphics[width= 7.8 cm]{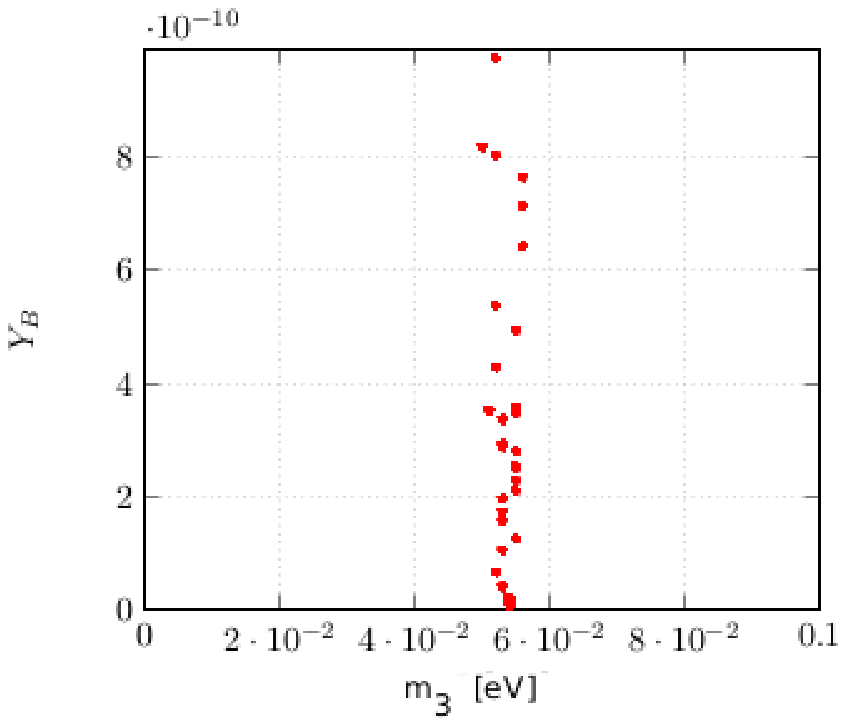}}
\caption{Scatter plot of the lightest neutrino mass $ m_{3} $ against the baryon asymmetry of the Universe with inverted hierarchy, non-unitarity case in one flavor leptogenesis regime.}
\end{figure}
\begin{figure}[tbh]
\centerline{\begin{subfigure}[]{\includegraphics[width= 6.5 cm]{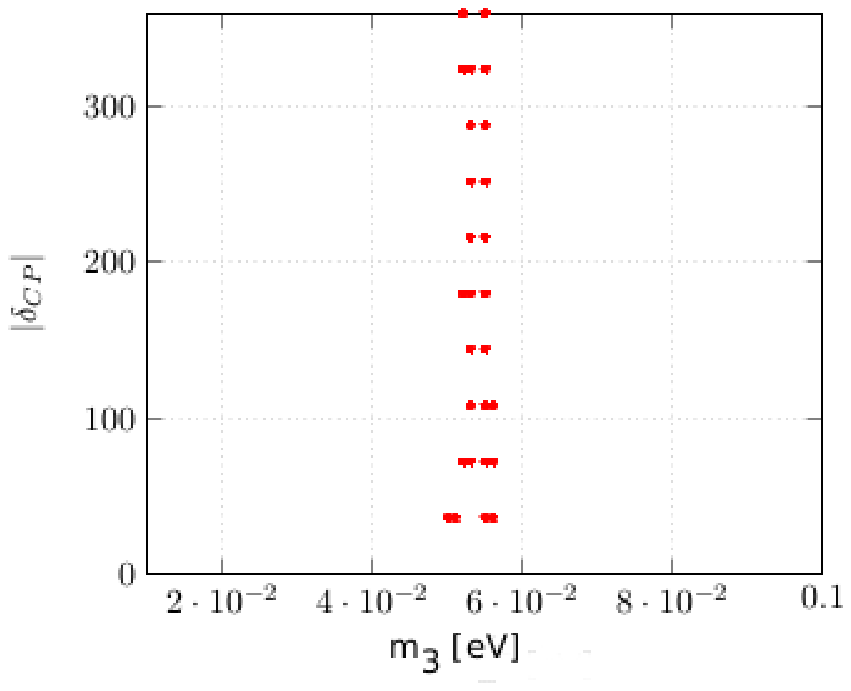}}\end{subfigure}
\begin{subfigure}[]{\includegraphics[width= 6.5 cm]{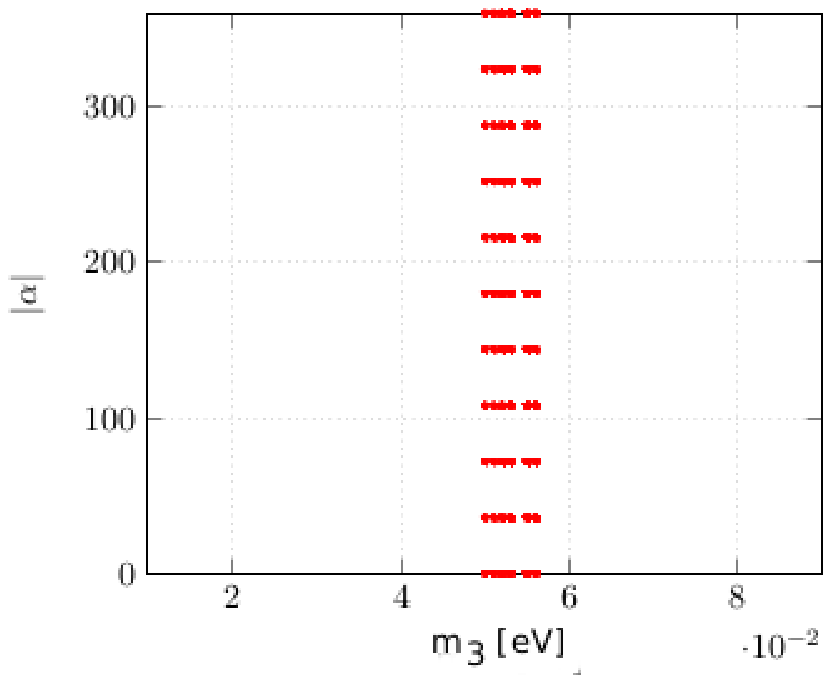}}\end{subfigure}}
\caption{Variation of lightest neutrino mass $ m_{3} $ against Dirac CP phase $ \delta_{CP} $ and  Majorana phase $ \alpha $ in case of IH, non-unitarity case in one flavor leptogenesis regime. The values of $ m_{3} $ [eV] in Fig. b along the x-axis is multiplied by $ 10^{-2}$.}
\end{figure}
\begin{figure}[tbh]
\centerline{\begin{subfigure}[]{\includegraphics[width= 6.5 cm]{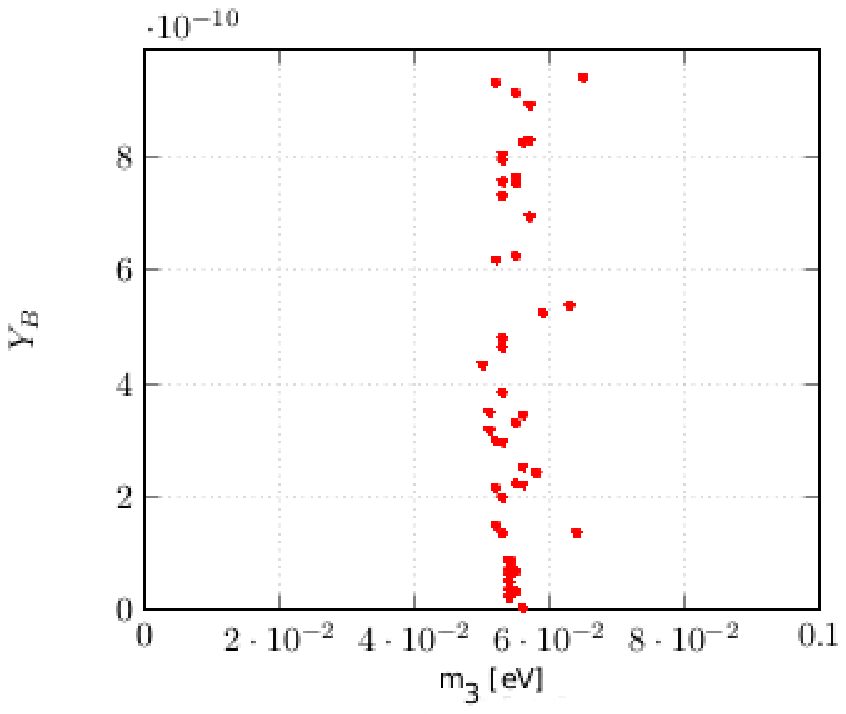}}\end{subfigure}
\begin{subfigure}[]{\includegraphics[width= 6.5 cm]{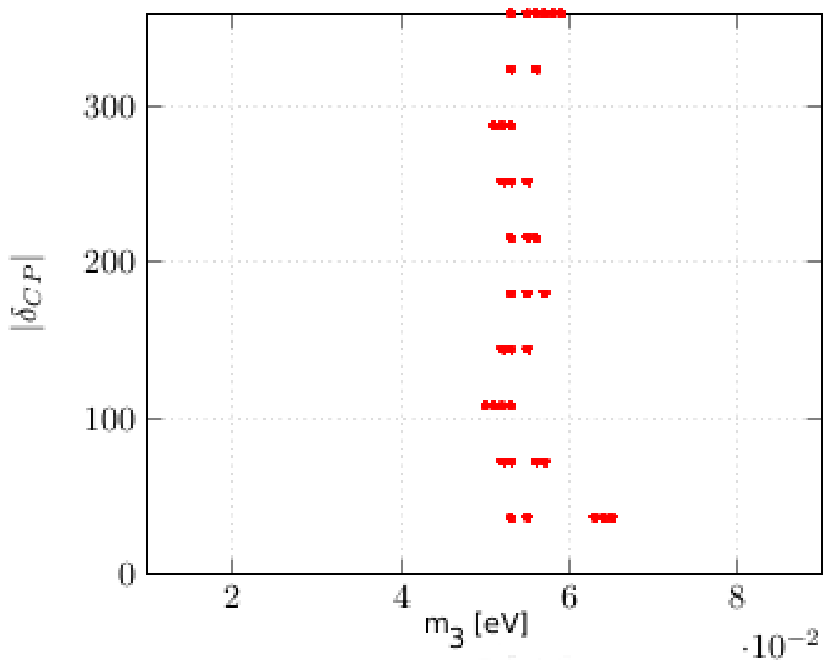}}\end{subfigure}}
\caption{Scatter plot of the lightest neutrino mass against the baryon asymmetry
of the Universe and Dirac CP phase $ \delta_{CP} $ with Inverted hierarchy, unitarity case in one flavor leptogenesis regime. For $ Y_{B} $ in the order $ 10^{-10} $ , the lightest $ \nu $ mass $ m_{3} $ is concentrated in the region 0.053eV to 0.062 eV. The values of $ m_{3} $ [eV] in Fig. b along the x-axis is multiplied by $ 10^{-2}$.}
\end{figure}

\begin{figure}[tbh]
\centerline{\begin{subfigure}[]{\includegraphics[width= 6.4 cm]{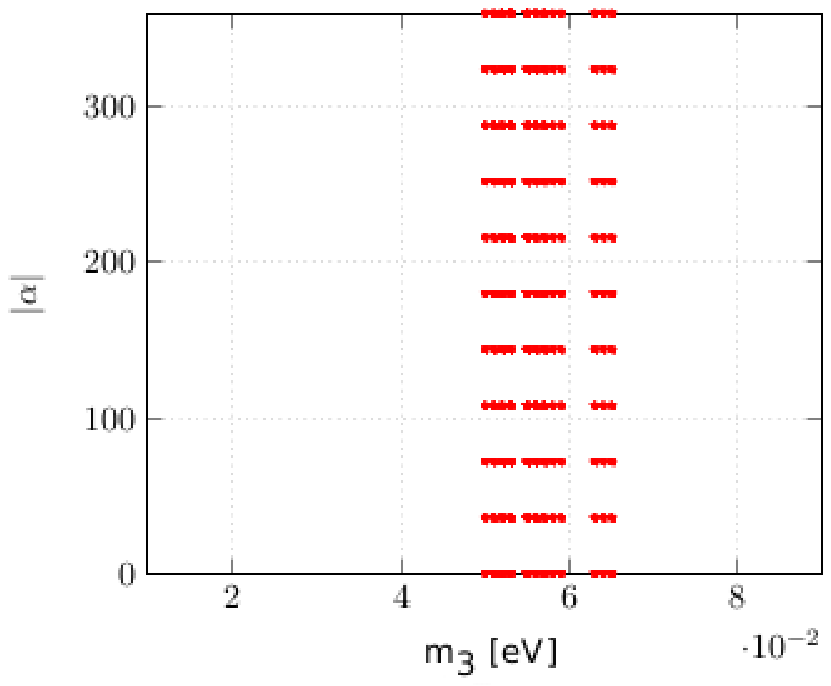}}\end{subfigure}
\begin{subfigure}[]{\includegraphics[width= 6.5 cm]{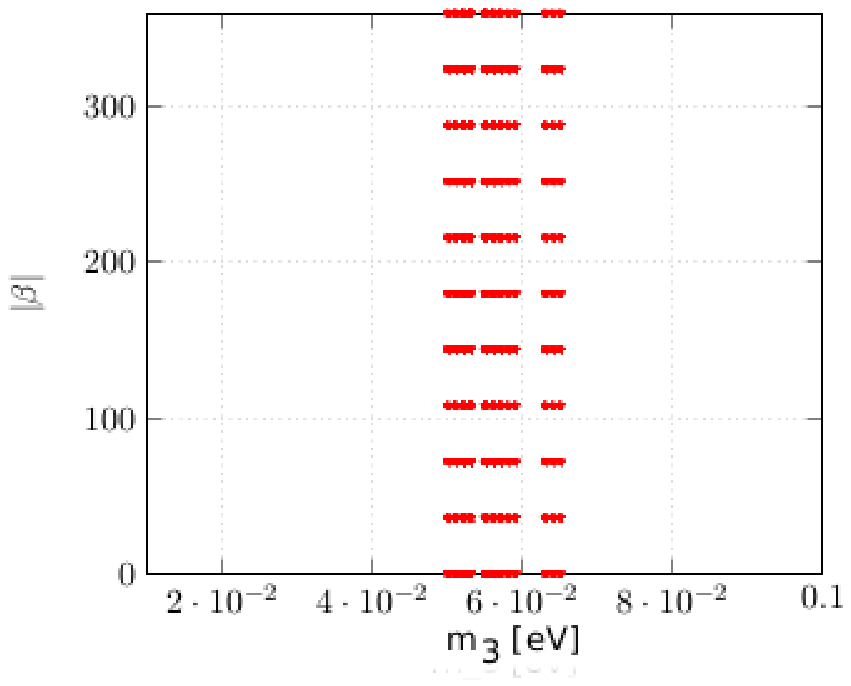}}\end{subfigure}}
\caption{Variation of lightest neutrino mass $ m_{3} $ with Majorana phase $ \alpha $ and $ \beta $ for IH, Unitarity texture of $ U_{PMNS} $ in one flavor leptogenesis regime. The values of $ m_{3} $ [eV] in Fig. a along the x-axis is multiplied by $ 10^{-2}$. }
\end{figure}
Figure 6 shows the scattered plot of lightest neutrino mass $ m_{3} $ against Dirac CPV phase $ \delta_{CP} $ and Majorana phase $ \alpha$ in IH of neutrino mass, non-unitarity case of one flavor leptogenesis regime. Figure 7 shows the variation of the lightest neutrino mass against the baryon asymmetry
of the Universe and Dirac CP phase $ \delta_{CP} $ with Inverted hierarchy, unitarity case in one flavor leptogenesis regime. For $ Y_{B} $ in the order $ 10^{-10} $ , the lightest $ \nu $ mass $ m_{3} $ is concentrated in the region 0.053 eV to 0.062 eV. In Fig. 8 we have shown the scatter plot of $ m_{3} $ with Majorana phase $ \alpha $ and $ \beta $ for IH, Unitarity texture of $ U_{PMNS} $ in one flavor leptogenesis regime. 
\par 
In Fig. 9, we have shown the effect of non-unitarity on probability $P(\nu_{\mu}\rightarrow \nu_{e}) $ for a particular case of Long Baseline Neutrino Experiments (DUNE, FNAL, USA). We have used the value of baseline to be L = 1300 Km and values of other oscillation parameters are taken from latest global fit \cite{Fg}. It can be seen that non-unitarity affects the probability, which means that its effect could be studied in neutrino oscillation experiments provided we reach the required precision level.

\begin{figure}[tbh]
\centerline{\includegraphics[width= 9.8 cm, height = 8.8 cm ]{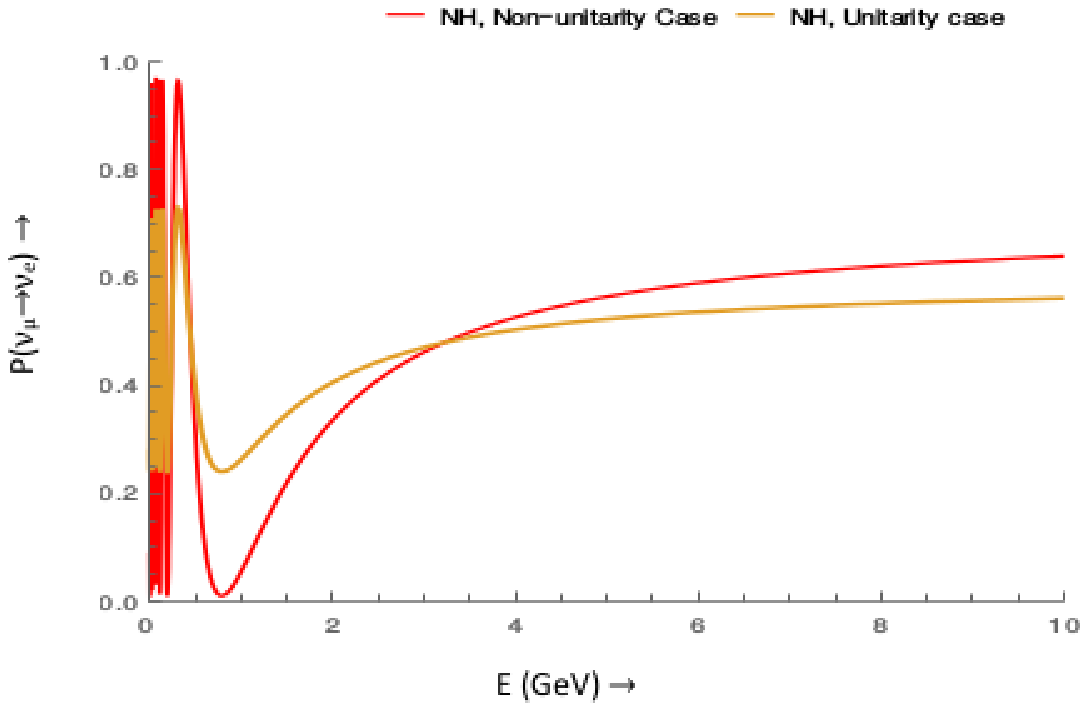}}
\caption{Variation of $ P(\nu_{\mu}\rightarrow \nu_{e}) $ against energy E in Long Baseline Neutrino Experiments with value of Dirac CP phase, $ \delta_{CP} = 1.45 \pi$, equal to the best fit value \cite{Fg}. The yellow (red) curve in the figure corresponds to unitarity (non-unitarity) of PMNS matrix. }
\end{figure}

\section{Analysis and Discussion}
We present here some comments, that reflect the main inference of this work. From the results presented in Table II we see that in all the four cases, there exist values of lightest neutrino mass, satisfying the constraint, $ \sum_{i} m (\nu_{i}) < 0.23 \text{eV}$ and those of present day baryon asymmetry of the Universe. The analysis of our results can be summarized as follows:
\newline
$\bullet$ One Flavor Leptogenesis: The value of lightest neutrino mass shifts to higher value in IH case, as compared to NH. Non-unitarity effects decrease the value of lightest neutrino mass in both NH and IH and the diminishing effect is more severe in NH.
\newline
$ \bullet $ We also found that non-unitarity affects the probability of $ \nu $ oscillation $ P(\nu_{\mu}\rightarrow \nu_{e}) $.
\newline
$ \bullet $ The predicted values of CP-violating phases, $ \delta_{CP} $ and Majorana phases $\alpha$ and $\beta$ are $36^{\circ}$, $72^{\circ}$, $108^{\circ}$, $144^{\circ}$, $180^{\circ}$, $216^{\circ}$, $252^{\circ}$, $288^{\circ}$, $324^{\circ}$, $360^{\circ}$.
\section{Conclusion}
To conclude, in this work, we have considered the possibility that the neutrino mixing matrix (considering charged lepton mass matrix to be diagonal), $ U_{PMNS} $ could be non-unitary, and then calculated the limits on non-unitary parameters $ \eta_{\mu e} $, $ \eta_{\tau e} $ and $ \eta_{\tau \mu} $ (see Table I) from latest constraints on branching ratios of cLFV decays. It is well known that in usual type I see-saw mechanism, mixing of left and right handed neutrinos may lead to non-unitarity but it has been found that \cite{pil} its effect is not significant for processes like lepton flavor violation and neutrino oscillation. Therefore we consider here a model (see Ref. \cite{Wern}) where see-saw is extended by an additional singlet (very light) which although induces non-unitarity of the $U_{PMNS}$ matrix, it leaves formula for see-saw mechanism unmodified. This non-unitarity however may affect leptogenesis. Baryogenesis through leptogenesis is believed to be responsible for producing the matter-antimatter asymmetry present in the present day universe, which can be expressed through parameter $ Y_{B} $ (baryon to photon ratio). We then analysed how the non-unitarity of $ U_{PMNS} $ can affect leptogenesis, and hence calculated the values of lightest $ \nu $ mass, Dirac CPV phase $ \delta_{CP} $ and Majorana phases $ \alpha $ and $ \beta $, such that $ Y_{B} $ lies in the present day constraints ($5.8\times 10^{-10} < Y_{B} < 6.6 \times 10^{-10} $). This was done using type I see saw mechanisms for producing light neutrino masses.
\par 
Above analysis was done for different cases $ - $ NH neutrino masses, unitary $ U_{PMNS} $;  NH neutrino masses, non-unitary $ U_{PMNS} $; IH neutrino masses, unitary $ U_{PMNS} $; IH neutrino masses, non-unitary $ U_{PMNS} $. We discussed these issues for unflavored leptogenesis regimes, for which $ M_{1} \geq 10^{12} $ GeV, where $ M_{1} $ is the lightest of the three heavy right-handed Majorana neutrinos, whose out of equilibrium decays produces lepton asymmetry (which in turn can be converted to BAU).

\begin{table}[phtb]
\begin{center}
\begin{tabular}{|c|c|c|}
 \hline
Case & $m_{lightest} $  & One Flavor\\
 \colrule
$\text{NH, non-unitarity}$ & $ 0.0018 \hspace{.1cm} \text{eV\hspace{.1cm} to}\hspace{.1cm} 0.0023 \hspace{.1cm}\text{eV} $ & $ \checkmark $\\
 
$\text{NH, unitarity}$& $ 0.048 \hspace{.1cm} \text{eV\hspace{.1cm} to}\hspace{.1cm} 0.056 \hspace{.1cm}\text{eV} $ & $ \checkmark $\\

$\text{IH, non-unitarity}$ & $ 0.05 \hspace{.1cm} \text{eV\hspace{.1cm} to}\hspace{.1cm} 0.054 \hspace{.1cm}\text{eV} $ & $ \checkmark $\\
 
$\text{IH, unitarity}$& $ 0.053 \hspace{.1cm} \text{eV\hspace{.1cm} to}\hspace{.1cm} 0.062 \hspace{.1cm}\text{eV} $ & $ \checkmark $\\
\hline
\end{tabular} \label{ta1}
\end{center}
\caption{Our calculated results for $ m_{lightest} $  with inverted hierarchy, normal hierarchy and one flavor leptogenesis. The symbol $ \checkmark $ ($ \times $) is used when $ Y_{B} $  is within (not within) updated BAU range.}
\end{table}
\begin{table}[htb]
\begin{center}
\begin{tabular}{|c|} \toprule
\hline
$ \delta_{CP} $, $\alpha$, $ \beta $\\
 \colrule
$36^{\circ}$, $72^{\circ}$, $108^{\circ}$,  \\ $144^{\circ}$, $180^{\circ}$, $216^{\circ}$,  \\ 
$252^{\circ}$, $288^{\circ}$, $324^{\circ}$, \\
$360^{\circ}$ \\
\hline
\end{tabular} \label{ta1}
\end{center}
\caption{The results for Dirac CPV phase $ \delta_{CP} $ and two Majorana phases $ \alpha $, $ \beta $  of all the four cases mentioned above in one flavor leptogenesis regime in this work.}
\end{table}

In this work, we have calculated new limits on non-unitarity parameters using latest bounds on cLFV decays and thus predicted values of lightest neutrino mass, (Table II) for both the hierarchies, which is still unknown experimentally. We also have predicted values of CPV phase $ - $ $ \delta_{CP} $ (Dirac phase) and $ \alpha $ and $ \beta $ (Majorana phases), which are also unknown so far (Table III). Though Majorana phases do not affect neutrino oscillation probability, they may affect neutrino mass measurements in $ 0\nu\beta\beta $ experiments. Hence the results in this work are important, keeping in view that in future, experiments will be endeavoring to measure the values of absolute value of neutrino mass, and CP-violating phase $ \delta_{CP} $ and $ \alpha, \beta $ (Majorana phases). Future measurements related to Dirac CPV phase in neutrino experiments will validate or contradict some of the results presented here. Our analysis in this work only provides a benchmark for consistent works affiliated to model building .
\section*{Acknowledgments}

GG and KB would like to thank Debasish Borah for fruitful discussions, and suggestions. GG would like to thank UGC, Government of  India, for providing RFSMS fellowship to her, during which this work was done. GG also acknowledges the fruitful discussions carried out on the constraints of the sum of the absolute neutrino mass scale at the XXII DAE BRNS High energy physics symposium, 12 $-$ 16 December, 2016, Delhi University, India.

\section*{References}

\end{document}